# Achieving Diverse and Monoallelic Olfactory Receptor Selection Through Dual-Objective Optimization Design

Xiao-Jun Tian[1,†], Hang Zhang[2†,], Jens Sannerud[3], and Jianhua Xing[1,*]

[1]Department of Computational and Systems Biology, School of Medicine, University of Pittsburgh, Pittsburgh, PA, 15260, USA

[2]Genetics, Bioinformatics, and Computational Biology Program, Virginia Polytechnic Institute and State University, Blacksburg, VA, 24060, USA

[3]The TECBio Research Experiences for Undergraduates Program, University of Pittsburg, Pittsburgh, PA, 15260, USA

[†]The two contributed equally

*To whom correspondence should be sent: xing1@pitt.edu

Running title: Olfactory Receptor Selection

**Abstract** Multiple-objective optimization is common in biological systems. In the mammalian olfactory system, each sensory neuron stochastically expresses only one out of up to thousands of olfactory receptor (OR) gene alleles; at organism level the types of expressed ORs need to be maximized. Existing models focus only on monoallele activation, and cannot explain recent observations in mutants, especially the reduced global diversity of expressed ORs in G9a/GLP knockouts. In this work we integrated existing information on OR expression, and constructed a comprehensive model that has all its components based on physical interactions. Analyzing the model reveals an evolutionarily optimized three-layer regulation mechanism, which includes zonal segregation, epigenetic barrier crossing coupled to a negative feedback loop that mechanistically differs from previous theoretical proposals, and a previously unidentified enhancer competition step. This model not only recapitulates monoallelic OR expression, but also elucidates how the olfactory system maximizes and maintains the diversity of OR expression, and has multiple predictions validated by existing experimental results. Through making analogy to a physical system with thermally activated barrier crossing and comparative reverse engineering analyses, the study reveals that the olfactory receptor selection system is optimally designed, and particularly underscores cooperativity and synergy as a general design principle for multi-objective optimization in biology.

**Keywords:** Bifunctional LSD1 / Diversity / Dual-objective optimization / Enhancer competition / Epigenetic modeling / Cooperativity /Ratchet/ Barrier-crossing

**Significance Statement:** For sensitive smell detection, each mammalian olfactory sensory neurons need to express stochastically only one allele of one out of possibly more than one thousand types of olfactory receptors. The mechanism for this mono-allelelic expression remains as one of the biggest unresolved questions for decades. Using mathematical modeling and computer simulations, we identified a three-layer regulation mechanism the olfactory system adopts to achieve single allelic expression and several other biological requirements such as maximizing the overall diversity of expressed olfactory receptors. The revealed mechanism provides insight for formulating biological processes as multiple-objective optimization problems.

# Introduction

For an engineer, successful design of a new product needs to meet multiple objectives such as maximizing targeted mechanical performance and minimizing the cost. Some of these objectives are incompatible, thus trade-offs are necessary. Similarly, living organisms are also constantly under selection pressure to maximize their fitness to the environment through optimizing multiple objectives such as growth rate and resistance to environmental fluctuations. A central task for systems biology is to unravel the corresponding mechanisms, or the design principles ultimately determined by evolution (1, 2), especially how a system prioritizes the multiple objectives and makes necessary compromises.

One example of multi-objective optimization is from the olfactory system. Olfaction, or the sense of smell, is essential for the survival and reproduction of an organism. Thus, most species have evolved a highly sensitive olfactory system. A major functional unit of the mammalian olfactory system is the main olfactory epithelium where up to millions of olfactory sensory neurons (OSNs) reside. These OSNs sense odorant molecules through transmembrane olfactory receptors (ORs), and transmit electric signals to the brain. OR genes are the largest gene superfamily in vertebrates. There are ~60 OR genes in drosophilas, 100-200 in fish, ~1,300 (including ~20% pseudogenes, *i.e.*, dysfunctional genes that have lost protein-coding ability) in mice and ~ 900 (including ~63% pseudogenes) in humans (3-7).

Proper function of the olfactory system imposes two basic requirements on the olfactory sensory neuron differentiation. First, in mammals, an individual OSN only stochastically expresses one type of functional OR, or more precisely one allele of the gene (6, 8-10). This monoallelic expression of OR proteins with rare violations has also been shown in other organisms such as catfish and zebrafish (11, 12). Expression of more than one type of OR would lead to improper stimulation and wiring of the olfactory system and thus misinterpretation of chemical signals (13). Second, each allele is expressed with approximately equal frequency in the neuron population (9, 13). Such diversity of OR expression maximizes the capacity of olfaction. Both monoallelic OR expression in a single neuron and maximal diversity of OR expression in the neuron population are essential for specificity and sensitivity of olfactory sensing.

The above observations raise one of the most intriguing puzzles in neurobiology that remains elusive after several decades of intensive investigations: how can both monoallelic and diverse expression of OR be ensured at the same time? Previously proposed molecular mechanisms focus only on the requirement of monoallelic OR expression (14-16), and are insufficient to explain a large amount of observations with various mutants.

By comparison, a key conceptual advance of the present study lies in the recognition that OSNs have evolved an optimal strategy for olfactory receptor activation as a dual-objective design problem with the following specific requirements. Before differentiation, all OR genes should remain transcriptionally silent. Within a biologically relevant period of time (5-10 days for mice) one allele is stochastically selected to become transcriptionally active and the error rate of multi-allele activation should be minimized. Furthermore, each gene has approximately equal probability of being activated so that the diversity of activated OR genes is maximized at the neuron population level. If a pseudogene is selected, it should be recognized and reselected until a functional allele is chosen. After differentiation the selected allele should be kept transcriptionally active while others remain inactive for the life-time of an OSN (about 100 days for mice). In the remaining parts of the paper, we will demonstrate that recognizing the dual-objective optimization requirement turns out to be essential for unraveling a robust three-layer regulation molecular mechanism, which both predicts existing observations and suggests new experiments.

## Results

**Mathematical formulation of OR activation is based on available experimental information.**

OSNs expressing different subsets of ORs topologically segregate into circumscribed zones. For example, zone 1 of the mouse main olfactory epithelium contains OSNs that express a subset of 150 OR alleles (4). Within each zone, the OR alleles in the corresponding subset are expressed with nearly equal probability (9, 17, 18). Similar segregated distribution has been found in zebrafish (19). Zonal segregation reduces the number of OR alleles competing for single allele expression from thousands to hundreds within a zone.

Recent studies revealed that an active OR allele in mice changes its epigenetic signature from H3K9me3, a covalent histone mark typically repressing gene transcription, to H3K4me3, a mark typically activating gene transcription, and this change is likely conserved in mammals (20). Similar epigenetic regulation was reported in zebrafish and *Drosophila* (21, 22). Furthermore, disruption of either histone methyltransferases or demethylases leads to violations of the rule of one-allele-activation and/or loss of diversity (22-24). Together with the observation that during OSN differentiation a histone demethylase LSD1 is transiently expressed, the above results suggest a competition among OR alleles for the H3K9me3-to-H3K4me3 transition (see Fig. 1A) (23).

Furthermore, it is well recognized that a feedback loop elicited by expression of the chosen functional OR gene maintains the selection and inhibits further activation of other OR genes (14-16, 23, 25-28). Recent studies reveal that expression of the winning allele causes endoplasmic reticulum stress and expression of enzyme Adcy3, which then down-regulates LSD1, leading to an epigenetic trap that stabilizes the OR choice (23).

Based on the above available information, we modeled a cell with 100 alleles to recapitulate the selection process within a single zone of olfactory epithelium, and formulated the following mathematical model for the OR activation problem within one cell as illustrated in Fig. 1. Throughout this paper for simplicity of presentation we treat the OR genes within a cell as a number of individual alleles. Each OR allele consists of a linear array of $N = 41$ nucleosomes, and each nucleosome can bear repressive H3K9 (R), no (E), or active H3K4 (A) methylations. Transition between these states is governed by enzyme concentration dependent rates. Specifically, demethylation steps R→E and A→E can take place either through stochastic exchange between nucleosome histones and the reservoir of unmarked histones with a turnover rate constant *days*, or through demethylation reactions with rates proportional to concentration of the catalyzing enzyme LSD1, which catalyzes both H3K4 and H3K9 demethylation. To maintain stable collective epigenetic state of an allele, previous studies reveal that the methylation state change on a nucleosome needs to be influenced by the methylation states of other nucleosomes beyond immediate neighbors (29, 30). Therefore we set the methylation rate constants $k_1$ and $k_2$ as functions of methylation states of other nucleosomes: $k_1$ ($k_2$) is promoted by H3K4 (H3K9) methylation in other nucleosomes, and the influence decreases with the nucleosome spatial separation. More model details are given in the Method section.

We propagated the nucleosome methylation states using stochastic Gillespie simulations, and simultaneously updated the levels of the expressed OR protein, Adcy3, and LSD1 by solving deterministic rate equations shown in Fig. 1B. We assumed that the gene is only epigenetically active when the fraction of nucleosomes bearing active marks, $\lambda$, is larger than a threshold $\lambda_\theta$.

**Low Noise and lack of demethylases kinetically freeze allele epigenetic state before and after differentiation**

We first examined the model under conditions prior to and after OSN differentiation when the LSD1 level is low. In this case, cooperation among nucleosomes biases them to have the same histone marks. This cooperation leads to collective epigenetic state dominated by either repressive or active marks (Fig. 2A), which is destabilized by removal of existing methylation marks on the nucleosomes, such as that induced by increasing LSD1 concentration (Fig. 2A-B) or the level of system noise due to stochastic histone turnover (Fig. S1A-B). These results are consistent with previous studies (29, 30). In other words, prior to and after differentiation, maintaining high levels of methyltransferases and low levels of demethylases forces an allele to be kinetically trapped at one of the two possible epigenetic states throughout the life time of an OSN, analogous to a system trapped in a double-well shaped potential with a very high barrier (see Fig. 2C). The above mechanism is confirmed with additional simulations through scanning 256 sets of parameters (Fig. S1C-D). In general maintaining stable epigenetic states requires that the methylation rates are much faster than the demethylation rates, and comparable propensity of adding both active and repressive marks, $i.e.$, $k_1/k_{-1} \sim k_2/k_{-2}$. The latter requirement can be relaxed when one or both of the demethylation rates are very low, then larger concentration fluctuations of the methyltransferases are allowed.

**Elevation of bifunctional demethylase level leads to a barrier-crossing like dynamics**

Next we analyzed the OSN differentiation process with bifunctional LSD1. As shown in Fig. 3A, after elevation of the LSD1 concentration at time 0, the OR alleles remain as repressive mark dominated, until one allele becomes active mark dominated, which leads to the corresponding OR expression and subsequent Adcy3 expression. Adcy3 down regulates LSD1, then the system maintains at a steady state with one OR allele active and the remaining ones inactive. Notice that the inactive alleles remain H3K9me3 dominated throughout the time. Due to stochasticity of the histone modification process, sampling over 1000 cells gives a broad distribution of $T_1$, the time of having the first allele epigenetically active, ranging from a few to 20 or more days and roughly centered around day 8 (Fig. 3B). Throughout their lifespan most of the OSNs only have one allele epigenetically activated, while a small fraction has two and rarely 3 alleles epigenetically activated (Fig. 3C), consistent with the functional requirements and experimental observations.

Close examination of the simulated trajectories reveals a simple mechanistic explanation for the monoallelic activation. Starting with the repressive mark dominated state, transient increase of LSD1 after initiation of OSN differentiation demethylates nucleosomes, and allows changing of methylation states in the nucleosomes. As a consequence, small patches of H3K4me3 nucleosomes may form, but are flanked by extended regions of H3K9me3 nucleosomes. Such H3K4me3 patches are unlikely to expand because of the cooperativity of methylation among nucleosomes and the dominance of H3K9me3 marks at the current stage. Nevertheless, when an H3K4me3 patch reaches a critical size -- as a rare event, it is able to propagate spontaneously and generate an epigenetic conversion of the OR gene into the H3K4me3 dominated state. That is, LSD1 increase resembles lowering the transition barrier between the double-well shaped potential shown in the previous section, and allows rare transition to happen (Fig. 3D). Once one allele converts to the H3K4me3 dominated state, and triggers the negative feedback loop to remove LSD1, the system is kinetically trapped again with high "transition barrier". The converted allele is kept active with H3K4me3 marks, while the remaining alleles bear repressive H3K9me3 marks. A prominent feature of this barrier-crossing-like dynamics is that throughout the process the probability of having an allele with hybrid pattern of epigenetic marks is low, and most alleles only fluctuate around the H3K9me3 dominated state (see Movie S1).

Based on the above analogy to a double-well potential, we reasoned that increasing the LSD1 concentration facilitates epigenetic state transitions. Indeed simulation results show that upon increasing the LSD1 concentration, $<T_1>$, the average of $T_1$, decreases (Fig. 3E), but the fraction of cells with multi-allele activation increases (Fig. 3F). Lyons et al. also observed fewer mature OSNs in mice with reduced LSD1 (23), as predicted in Fig. 3F. Therefore, for a given number of alleles in the OR pool, an optimal LSD1 concentration may evolve to compromise the requirements of single-allele activation and efficient OSN differentiation.

Next we asked how the number of permitted alleles affects the ratio of cell with single allele epigenetic activation (Fig. S2). The ratio first increases since a cell with more alleles has higher probability to have at least one allele epigenetically activated during the differentiation period. Then it decreases after a peak value since the probability of having more than one allele activated also increases with the number of alleles per zone. While the exact position of the peak depends on model parameters, the model results predict that the number of OR genes within a zone is under selective pressure.

**Comparative model studies reveal that Nature chooses the simplest and robust design of feedback regulation.**

The fact that the feedback regulates a bifunctional demethylase, LSD1, seems both counter-intuitive and inefficient, since the enzyme removes both the repressive and active methylation marks, with the latter being what added to an active allele. Theoretically the feedback could act on any one or any combination of the four groups of enzymatic reactions (Fig. S1C). Therefore we simulated all the 10 cases that the feedback regulates one or two of the reaction rates. All these cases have the same set of parameters for cells after activating the feedback, and they differ only on value(s) of one or two rates prior to feedback taking effect. By scanning each combination of parameter pairs over a $7 \times 7$ grid and performing 500 independent simulations for each parameter set, indeed the case of negative feedback on both of the two demethylation rate constants, *i.e.*, a bifunctional LSD1, leads to the highest monoalleleic activation ratio (Fig. S3). A less robust scheme requires regulating the H3K9 demethylase and H3K4 methyltransferase oppositely at the same time. Both of these schemes modulate the effective transition "barrier" without necessarily changing the relative stability of the two collective epigenetic states.

To further understand the critical role of the bifunctional LSD1, we examined one of the above hypothetical cases that unifunctional LSD1 only catalyzes H3K9 demethylation. In this scenario, the system proceeds with a ratchet-like dynamics (31), and has a much higher ratio of multi-allele epigenetic activation as well as much higher percentage of alleles trapped in the hybrid epigenetic state for an extended period of time after the LSD1 level is reduced (Fig. S4A-E). An allele in a hybrid epigenetic state has some nucleosomes bearing H3K9me3 and others bearing H3K4me3. Such hybrid state is not normally present in stable cell phenotypes, and extended period of existence in this state is likely detrimental for a cell since histone marks can affect higher-order chromatin structures and gene activities (32).

To understand why the effective two-state barrier-crossing dynamics is advantageous over the multi-state ratchet-like dynamics on generating single allele activation, we performed further mathematical analysis based on the following reasoning. In the OR system a number of alleles convert their epigenetic state independently and stochastically under an elevated LSD1 concentration. Let us denote the activation time separation between the first two converted alleles as $\tau$. Then from an engineering perspective, a better design to achieve single-allele activation is the one with a larger $\tau$, which means that the two activation events are better separated temporally, and thus more time for the first allele to elicit the feedback loop and prevent activation of another allele. Therefore we performed mathematically controlled comparison among a set of n-state Markov chain models shown in Fig. S4F. Consider two alleles transiting independently from the repressive mark dominated state to the active mark dominated state through various numbers of intermediate states, but with the same mean first arrival time. Figure S4G shows that the two-state model has an exponentially shaped first-arrival-time distribution $f_2$, while those with $(n - 2 > 0)$ intermediate states have peaked ones that at large $t$ decrease faster with increasing $n$. One can randomly draw two points from a distribution, corresponding to the stochastic activation events of the two independent alleles. Clearly the temporal separation of the two points, $\tau$, is likely to be larger if they are drawn from a broader $f$ corresponding to smaller $n$. Indeed Fig. S4H shows that the distribution of $\tau$ has longer tail for smaller $n$. That is, a design with the two-state dynamics is better than that with the multi-state dynamics.

**Epigenetic competition model predicts zebrafish but not mouse experiments on inhibiting methyltransferases/demethylases**

With the above constructed model, in this section we compare model predictions to experimental observations. In the illustrative double-well potential shown in Fig. 3D, lifting the left well allows easier transition to the right well thus higher probability of multiple allele activation, while elevating the barrier height leads inhibition of allele activation. Experimentally, reducing the left well depth can be realized by reducing the enzymatic activity of H3K9 methyltransferases, G9a and GLP. Indeed, partially inhibiting the enzymatic activities of G9a/GLP ($E^R_{K9M}$) leads to increased number of cells coexpressing multiple ORs (Fig. 4A), which is confirmed in zebrafish (22).

Similarly decreasing LSD1 concentration corresponds to increasing the barrier height. The simulation results in Fig. 4A predict that reducing the LSD1 concentration ($LSD1^R$) impedes OR activation, which can be partially restored by decreasing the enzymatic activities of H3K9 methyltransferase ($LSD1^R/E^R_{K9M}$). The prediction has been confirmed in mice (24).

However, the model so far predicts a phenomenon different from experimental observation, if H3K9 methyltransferase activity further decreases. As the left well in Fig. 3D lifts further the system increasingly resembles that in Fig. S4E. In other words, the epigenetic activation process evolves from the barrier-crossing-like dynamics to the ratchet-like dynamics. The simulated results in Fig. S5A-B indeed predict that further reducing the level of H3K9 methyltransferase ($E^{RR}_{K9M}$) drives a majority of the OR alleles into hybrid methylation pattern during the differentiation process, and thus causes significant increase of multi-OR epigenetic activation (Fig. 4A). However, G9a/GLP double knockout (dKO) mice demonstrated elevated but still rare multi-OR coexpression compared to WT mice (24). Therefore the epigenetic conversion mechanism fails to predict the experimental results in mice.

**Competition for cooperatively bound enhancers further reduces coexpression of multi-allele ORs.**

To explain the G9a/GLP dKO mouse experiment, we noticed recent studies on enhancer-mediated regulation of OR expression (24, 33). Multiple regulatory genome sequences, *i.e.*, enhancers, bind to the promoters of active OR alleles, but not the silenced ones, and form a dense interaction network, possibly mediated by DNA and histone binding proteins such as transcription factor Bptf (33-36). Therefore these enhancers possibly act as *cis* elements during the OR selection process. Indeed disruption of the enhancer elements affects OR expression (34, 36). We hypothesize that for terrestrial vertebrates such as humans and mice, active expression of an OR allele requires both the gene bearing active epigenetic marks (H3K4me3) and co-localization of a sufficient number of enhancers to the allele. In the following, we will present the generalized model that incorporates enhancer binding, and show how it provides a backup selection mechanism to guarantee diverse and monoallelic activation when combined with the epigenetic selection discussed above.

Suppose $M$ enhancers are available for an OR genomic cluster with $L$ OR alleles (see Fig. 4B). Each enhancer can bind to the epigenetically active $l$-th OR allele with a free energy of binding $\varepsilon_l$, and can interact with any other enhancer bound to the same allele with energy $\zeta$. Enhancer binding to alleles with repressive marks is weak and can be neglected (33). Notice that enhancer binding to active alleles is cooperative: when two or more epigenetically active alleles compete for the enhancers, an enhancer preferentially binds to the allele that already has more enhancers bound since more enhancer-enhancer interactions can form. Consequently, enhancers collectively bind to and transcriptionally activate one allele at a given time; switching an enhancer from one allele to another is rare since it requires breaking many interactions. This mechanism inhibits multi-allelic expression in the rare cases in which more than one allele is epigenetically activated. We performed Gillespie simulations for the enhancer-allele binding/unbinding dynamics. Figure 4C-D give an example of a cell with two alleles becoming epigenetically active, but only one of them is transcriptionally active at a given time.

While the exact number of enhancers, $M$, is not known, for proper functioning of the backup selection mechanism, $M$ is likely smaller than the number needed for saturating an allele; otherwise, the superfluous enhancers could bind to other alleles and sabotage monoallelic expression. Indeed, it is experimentally observed that ectopically introduced multiple copies of a specific H enhancer increase the probability of multi-OR coexpression (34).

We then investigated the effect of enhancer competition between epigenetically active alleles. If the enhancers bind to the two alleles with equal strength, *i.e.*, $\zeta$ and $\varepsilon_l$ assume the same values for different alleles and enhancers, the enhancers jump stochastically and collectively between the two alleles, showing a two-state dynamics alike a particle moving in a symmetric double-well potential (left panel of Fig. 4D). The frequency of transitions depends on the actual binding strength and the number of enhancers. However, it is likely that the values of $\zeta$ and $\varepsilon_l$ are slightly allele-dependent due to differences in sequence or enhancer-promoter proximity. Then cooperative enhancer binding can amplify this difference by many folds. For example, suppose that there exists a free energy difference of enhancer-allele binding $\Delta\varepsilon = \varepsilon_1 - \varepsilon_2$ between allele 1 and allele 2. Then the free energy difference between allele1 bound with $M$ enhancers and allele 2 bound with $M$ enhancers is $M\Delta\varepsilon$, which can be significant due to the factor $M$. So the allele with stronger enhancer binding dominates transcriptionally alike a particle moving in an asymmetric double-well potential (middle and right panels of Fig. 4D).

The above model results lead to a surprising prediction on the OR expression pattern when the level of H3K9 methyltransferases is reduced. Compared to WT cells, the cells with $E_{K9M}^{R}$ tend to have more OR alleles being epigenetically active (Fig. 4E), as expected. However, except for a small group of OR genes becoming transcriptionally upregulated, most of them instead show decreased expression compared to those in the WT system (Fig. 4F). Further reduction of the enzyme level ($E_{K9M}^{RR}$) causes fewer OR alleles to be expressed, but each with higher expression level (Fig. 4F-G). This seemingly counterintuitive prediction has been confirmed experimentally (24).

Here we use a toy system to illustrate why the enhancer-mediated regulation, when combined with weak or no epigenetic selection, causes reduced diversity of OR expression (Fig. 4F). Suppose $L$ (= 4) OR alleles exist in a zone, and these alleles have strong (allele 1), medium (allele 2), and weak (alleles 3 and 4) binding strength to the enhancers, respectively (Fig. 4H). Existing experimental evidences suggest that the epigenetic activation step is stochastic and each allele has roughly equal probability $1/L$ to be chosen. For WT OSNs, most cells have only one epigenetically active allele, and the allele becomes transcriptionally active as well. Therefore the overall transcriptional probability of each allele in the zone is ~ ¼. On the other hand, with the H3K9 methyltransferase level reduced ( $E_{K9M}^{R}$ and $E_{K9M}^{RR}$, or G9a KO and G9a/GLP dKO experimentally), an OSN may have multiple epigenetically active OR alleles. For simplicity of argument let us assume that in a cell three alleles compete for enhancers. Since each allele has the same probability of becoming epigenetically active, there are 4 possible combinations with equal probability, (123), (124), (134) and (234). As an allele with stronger enhancer binding dominates transcription, one expects that the first 3 combinations mainly express allele 1, and the last one expresses allele 2. That is, the expression of allele 1 is upregulated while that of alleles 3 and 4 are down regulated. Similarly, with more epigenetically active alleles coexisting in individual OSNs, the alleles that bind enhancers more strongly secure greater chances of outcompeting other epigenetically active allele in an OSN and getting expressed, whereas the weaker alleles have little such chance. Consequently, the OR diversity in the OSN population diminishes.

In the above simulations we assumed that only the number of enhancers bound to an allele affects its transcription. It is possible that enhancers have certain OR gene specificity (26, 34, 36). Therefore we considered the alternative possibility that only one of the binding enhancers, say enhancer 1, is necessary for activating a given OR gene. Compared to the case with only enhancer 1 (Fig. S5C upper panel), with other enhancers being present enhancer 1 shows increased dwelling time of binding to allele 1 and this binding correlates with the overall collective binding state of enhancers (Fig. S5C lower panel). That is, the presence of other enhancers stabilizes the binding of the enhancer who actually affects the allele transcription, and the above results discussed in this section still hold in this case.

**Model studies predict multiple mechanisms of OR transcription switching**

The trajectories in Fig. 4D reveal that an OSN cell occasionally switches off an active OR allele and chooses another one. Such switching phenomena have been widely reported in the literature (18). In this section we summarize several scenarios of the switching behavior predicted by the model.

First, a pseudogene allele gets epigenetically and transcriptionally activated. The pseudogene fails to generate OR proteins to elicit the Adcy3 mediated feedback loop to reduce the LSD1 level, and thus permits another allele to be epigenetically activated and transcriptionally switched on. We observed most of the switching taking place in WT OSNs of this type (Fig. 5A).

Second, a functional allele gets epigenetically and transcriptionally activated, but the OR-Adcy3-LSD1 feedback fails to prevent another allele from becoming epigenetically activated. There is probability that the second allele wins over the first allele for the enhancers. If the second allele is functional, the cell switches its OR expression. If the second allele is a pseudogene (as exemplified in Fig. 5B), it becomes Scheme 1 and the cell re-enters the selection process until another functional allele becomes both epigenetically and transcriptionally activated. We observed this type of switching rarely in WT OSNs, but more in Adcy3 KO cells.

Compared to both scenario 1 and 2 that an allele remains epigenetically active even after switching off transcriptionally, an epigenetically activated allele may switch back to epigenetically inactivate state. The simulation results in Fig. 5C show a sequence of switching events. First a transcriptional switch takes place between two alleles (as in scenario 1). Then the newly activated allele switches off epigenetically and thus transcriptional, and the original allele switches back to be transcriptionally active. We observed this scheme only in Adcy3 KO mice where the feedback loop is disrupted so the sustained high level of LSD1 leads to collective removal of H3K4 methylation from the activated allele.

Not surprisingly, the switching frequency increases in Adcy3 KO OSNs compared to that in WT OSNs (Fig. 5D) since more cells have multiple epigenetically active alleles. Furthermore, the fraction of cells expressing pseudo ORs increases while that expressing functional ORs decreases in Adcy3 KO simulations (Fig. 5E). These predictions have been confirmed (23). Mechanistically in the Adcy3 KO system transcription of a functional allele does not inhibit further epigenetic activation of pseudogene alleles, and the latter then competes with the former for transcription.

Mechanistically, the model suggests two possible modes of switching OR expression. In the first mode, an allele converts from the active H3K4me3 epigenetic state back to the repressive H3K9me3 state. Experimental testing of this mechanism requires monitoring the histone modification state of one allele over time. In the second mode, the enhancers cooperatively change their binding from one allele to another one, with both being epigenetically active. The present model predicts that the genes showing upregulated expression in the G9a/GLP dKO mice, such as Olfr231, have slighter stronger interactions with the enhancers than the remaining genes do. Then an experimentally testable prediction is that in normal mice, OSNs that express one of these genes should have lower frequency of switching than those cells express other genes in the same zone do.

## Discussion

Monoallelic OR activation in olfactory sensory neurons is a decades-long puzzle in neurobiology. Recently several mathematical models have been formulated to examine various proposed mechanisms for explaining the phenomenon (14-16). Compared to these existing modeling studies, the present model integrated a number of key experimental observations available only recently. The model, while coarse-gained, has every of its components corresponding directly to an experimentally measurable quantity, which makes comparison to experimental results and prediction test transparent, as well as reveals some key design principles of the system.

**A sequential three-layer regulation mechanism controls single allele activation.**

A,major conceptual difference between the present model and others is that we emphasize the importance of treating OR activation as a multi-objective optimization problem. Our theoretical studies demonstrated that a series of selection processes functioning synergistically lead to diverse and single allele activation (Fig. 6). A subset of the available alleles is selected by the zonal segregation. Then they are randomly chosen to be epigenetically activated through transient elevation of bifunctional LSD1. Most of the cells only have one epigenetically active and thus transcriptional active allele. If more than one allele is epigenetically activated, they compete for a limited number of enhancers to be transcriptionally active, resulting in only one epigenetically and transcriptional active allele. If the activated allele is not a pseudogene, it triggers a feedback to prevent further epigenetic state change. Therefore, this coordinated three-layer regulation mechanism faithfully assures that only one OR allele stochastically selected with about equal probability and expressed in one OSN. Recent single cell sequence studies (37, 38) reveal more frequent violation of monoallelic OR expression in OSNs from newborns than those from adult mice. These observations are consistent with the present model since OSNs from the newborn have more dynamic chromosome structure, *i.e*., more enhancers accessible to an OR gene, than those of adults.

**The OR selection process is optimized to satisfy prioritized multi-objective requirements.**

Epigenetic activation leads to a large percentage of cells having single epigenetically active allele, and selects OR alleles with approximately equal probability. On the other hand, enhancer competition is more effective on ensuring single allele activation, but it also introduces strong bias towards allele selection.   Therefore, to achieve single allele activation as the top priority and maximize the diversity of expressed ORs at the same time, the OR selection system has evolved into a combined procedure. The epigenetic activation step is optimized with a bifunctional LSD1 to achieve maximal single allele activation. When multiple allele epigenetic activation does happen but with low probability, the enhancer competition in allele selection serves as the last "safeguard" without severely distorting the overall diversity of OR expression. Similarly our analysis reveals that other variables are also subject to the multi-objective optimization. For example, the LSD1 concentration may be optimized as a result of compromise between maximum single-allele activation and fast allele activation.

**Counter-intuitive bifunctionality of the LSD1 maximizes single allele epigenetic activation and minimizes the probability of hybrid state trapping.**

An intriguing feature of the OR selection system is that the selection is initialized then maintained through regulating the level of the bifunctional LSD1 that removes both repressive and active marks during the activation process. Our analysis shows that this bifunctionality leads to a barrier-crossing-like dynamics with high single allele epigenetic activation ratio and minimization of alleles trapped in hybrid epigenetic states. An alternative scheme such as the unidirectional LSD1 would lead to a number of OR genes in hybrid states, and it is hindered to relax the hybrid states back to the H3K9me3 dominated state after LSD1 reduction due to the negative feedback.

Therefore our model predicts that throughout the selection process a "tug-of-war" exists between adding and removing H3K9 and H3K4 methylations. This "tug-of-war" is analogous to that of ultrasensitive phosphorylation-dephosphorylation cycle observed in signal transduction networks (39), and works together with nucleosome crosstalks to generate the kinetic cooperativity during the epigenetic activation process. Furthermore it is necessary that the enzymatic activities of methyltransferases are in excess over that of demethylase, i.e., LSD1. Lyons *et al.* indeed observed that G9a/GLP at excessive concentration coexist with LSD1 during OSN differentiation (24).

A debate in the field is how the epigenetic race process is coupled to a feedback loop. Using the barrier-crossing analogy, the feedback can modulate either the relative stability of the two collective epigenetic states, as previous modeling studies emphasize (14, 15), or the transition barrier. Our comparative studies reveal that regulating the bifunctional LSD1, corresponding to varying the barrier height, is optimal. We also identify a number of less robust schemes corresponding to regulate both the barrier and the relative state stability.

**The model studies make multiple testable predictions.**

Our model makes multiple testable predictions. Table 1 summarizes our model predictions, experimental confirmations and suggested new experiments. Here we discuss a few of them in detail.

To reach the diversity change prediction in Fig. 4F & G, a key ingredient in the model is that the values of $\zeta$ and/or $\varepsilon_l$ are allele-dependent. The difference may come from DNA sequence, and it may be even less than the thermal energy $k_BT$, the product of Boltzmann's constant and temperature. However, this free energy difference can be significantly amplified by enhancer cooperative binding (Fig. 4D). This amplification explains why strong OR expression bias occurs in G9a/GLP dKO mice while Lyons *et al.* could not identify any significant differences between the promoters of the most upregulated ORs and the remaining ones in predicting the transcription-factor-binding-motifs (24). Another possible source of different OR-enhancer binding strength lies in the different distances between enhancers and alleles. Different allele-enhancer distances may require slight different DNA distortion to form the OR-enhancer binding complex, as implied by the observation that moving the H enhancer closer to MOR28 dramatically up-regulates its expression while down regulates other neighboring ORs (26). To further test this mechanism, one can replace an upregulated OR gene and its promoter by a down-regulated one, and test whether the latter becomes upregulated in a G9a/GLP dKO main olfactory epithelium. Another suggested experiment is to introduce enhancers ectopically to G9a KO mice (34), which should at least partially rescue the reduction of OR diversity if the model holds.

To test the prediction given in Fig. S5A&B, one may sort GFP$^+$ cells from OMP-IRES-GFP control mice and G9a/GLP dKO OMP-IRES-GFP mice (20), respectively, then perform CHIP-qPCR for selected silent OR genes. We expect that H3K9me3 dominates on silent OR alleles from the control mice, but H3K4 and H3K9 methylations mixed at various extent on silent OR alleles from the dKO mice (Fig. S5B). One can further measure the epigenetic pattern at different time points before and after differentiation to test the prediction that it takes long time for the alleles with mixed methylations to relax to a steady state distribution.

In summary, we have constructed and analyzed a comprehensive model that revealed a mechanism for achieving diverse and mono-allelic OR gene expression. A proper combination of mechanisms, but none of the individual one, can achieve the desired diverse and monoallelic OR expression. Given that multi-objective optimization is ubiquitous in biological systems, this synergetic and sequential application of different mechanisms is likely to be a general design principle on biological process regulation, and shed light on problems in other fields as well. This work aims at using a minimal model to reveal the essential elements that regulate the OR selection process. For example, chromatin structures in OSNs are highly dynamic to expose or sequester specific OR genes. Specific patterns of DNA methylation and other histone covalent modifications have been observed for OR promoters and enhancers. OR genes are not expressed with exact equal probability, and coordinated expression might exist (40). Furthermore, enhancer elements may also help on recruiting histone modification enzymes, leading to coupling between the two layers of regulation. Future studies will reveal these possible fine-tuning elements and address its implications in other processes of gene regulations.

## Materials and Methods:

Each OSN is modeled to have $N_p = 30$ pseudogene alleles and $N_f = 70$ functional OR alleles, with the only difference being that the product of the former does not elicit Adcy3 mediated feedback.

**Epigenetic dynamics:** For simplicity we treated step-wise methylations/demethylations on a nucleosome as single steps, and treated participating enzymes other than LSD1 implicitly. Denote methylation state of a nucleosome R, E, and A as $s = -1, 0, 1$, respectively. We set the methylation rate constants for an empty nucleosome $i$ as

$$k_1^i = k_1^0 \exp\left(\sum_{j \neq i} \frac{m}{|i-j|} \delta_{s_j,-1}\right), \quad k_2^i = k_2^0 \exp\left(\sum_{j \neq i} \frac{m}{|i-j|} \delta_{s_j,1}\right),$$

where the sum is over all other nucleosomes, and δ is a Kroneck-delta function. That is, each of the other nucleosomes influences the nucleosome to add the same mark of the latter, and the influence decreases with the nucleosome spatial separation. An insulating boundary is assumed, and three nucleosomes in the middle form a nucleation region with higher enzymatic rate constants than other nucleosomes have. We modeled $E^R_{K9M}$ and $E^{RR}_{K9M}$ by reducing the value of $k^0_2$ for the WT to 90% and 80%, respectively. Values of the model parameters can be found in Table S1.

Three nucleosomes located at the center of the nucleosome array form the nucleation region. Existence of this nucleation region reflects the observation that some DNA sequence specific molecular species, such as transcription factors and noncoding RNAs, help on recruiting histone modification enzymes. We also performed simulations without the nucleation region and the found no qualitative change of the mechanisms discussed in the present paper.

**Enhancer binding dynamics:** For simplicity we assumed that there is no free enhancer. This assumption is not essential for the present discussions and can be easily removed at the expense of a few additional parameters. Also we treated the enhancers equally, although generalization is straightforward when additional experimental information becomes available. An enhancer can jump from allele $i$ to $j$ with rate, $k_{i \to j} = v \exp[0.5(\varepsilon_i - \varepsilon_j + (M_i - 1 - M_j)\zeta)]$ to satisfy the detailed balance requirement, where $M_i$ and $M_j$ are the number of enhancers bound to allele $i$ and $j$ before the jump, respectively, and $\sum_i M_i = M$. We chose the factor 0.5 to satisfy the detailed balance requirement, i.e., $k_{i \to j}/k_{j \to i}$ equals to the Boltzmann factor corresponding to the system free energy after the transition divided by that prior to the transition. At each Gillespie simulation step, one of all possible enhancer binding changes is randomly selected. Since an allele with higher enhancer binding affinity dominates enhancer competition, for computational efficiency we only simulated enhancer dynamics explicitly for the results in Fig. 4D and Fig. S5. For other simulations in Fig. 4-5 we adopted a simplified procedure as schematically illustrated in Fig. 4H. That is, we stochastically ranked the enhancer binding affinities of the 100 alleles and let the one with highest enhancer binding affinity transcriptional active when more than one allele is epigenetically activated,

**Gene expression dynamics:** All gene expression is modeled by solving ordinary differential equations (Fig. 1B). For simulations with enhancer binding dynamics in Fig. 4-5, we multiplied to the first synthesis term of OR expression a Kroneck-delta function, which assumes 1 if the allele is epigenetically active and other alleles are epigenetically silent, or if its enhancer binding affinity is stronger than that of other epigenetically active alleles, and 0 otherwise. For Adcy3 KO simulations, $k_A$ is set to be 0. All concentrations are in reduced unit.

More details are provided in SI Materials and Methods.

## Acknowledgements

We thank Drs David Lyons, Andrew Chess, Jing Chen and Ken Kim for many helpful discussions. The research was supported by the U.S. National Science Foundation (DMS-1545771 and DMS-1462049). J.S. was supported by NSF/Department of Defense DBI-1263020, which funds the TECBio Research Experiences for Undergraduates (REU) program.

## Author Contributions

JX conceived the project, constructed the model and wrote the paper with input from XT and HZ. XT performed most simulations. HZ collected biological background information. HZ and JS also participated simulations. XT, HZ, JS, and JX analyzed the data.

## Conflict of Interest

The authors declare that they have no conflict of interest.

# Figure legends

**Figure 1. Mathematical model of the experimentally revealed regulatory system of olfactory receptor activation.** (A) Feedback regulated OR allele epigenetic activation. Each OSN contains $N_p$ ( = 30) pseudo OR alleles and $N_f$ ( = 70) functional OR alleles. Each allele is composed of a linear array of 41 nucleosomes. Each nucleosome bears active, no, or repressive mark, and a mark-bearing nucleosome facilitates an empty nucleosome to add the same mark in a distance dependent manner. Expression of an OR protein elicits a feedback to induce expression of enzyme Adcy3, which removes the demethylase LSD1. (B) The corresponding mathematical formulation. A nucleosome changes its covalent modification state stochastically with the indicated rate constants. The methylation rate constants $k_1$ and $k_2$ are influenced by nearby nucleosomes. Protein level changes are simulated by ordinary differential equations. $H(x)$ is a Heaviside function which assumes value 0 for x <0, and 1 otherwise. $\lambda_i$ is the fraction of active mark in allele i, while $\lambda_\theta$ is the cutoff fraction of nucleosomes with active marks so an allele is regarded as epigenetically activated.

**Figure 2. Low Noise and demethylation enzyme concentration kinetically freeze allele epigenetic state**. (A) Typical single allele trajectories of the fraction of nucleosomes with active marks under various constant concentrations of LSD1. (B) The fraction of alleles that maintain epigenetic state longer than 100 days under various constant concentrations of LSD1. The result was sampled over 1000 cells initially in the collective repressive mark dominated state. (C) Analogous double-well potential system with the barrier height inversely related to LSD1 concentration.

**Figure 3. Bifunctional LSD1 leads to barrier-crossing-like dynamics and ensures mono-allelic epigenetic activation.** (A) Typical trajectories of the fraction of nucleosomes with active marks on one allele for 100 alleles (represented by different colors) within a cell. The temporal change of LSD1 level (blue curve, in relative unit) is also indicated. (B) Distribution of $T_1$, the time observing the first epigenetically active allele (75% nucleosomes bearing active marks). Sampled over 1000 cells. (C) Fraction of cells with various numbers of epigenetically active alleles at day 100. (D) The analogous potential system during activation. (E) Dependence of the average of $T_1$ on the elevated LSD1 level ($[LSD1]_0$) during differentiation. (F) Dependence of the fraction of cells with various numbers of epigenetically active alleles at day 100 on $[LSD1]_0$. In all simulations a cell has 100 OR alleles, and at time 0 the LSD1 level is elevated 10 folds from its basal value to simulate the onset of differentiation.

**Figure 4. Competition for cooperatively bound enhancers further reduces co-expression of multi-allele ORs.** (A) Predicted fractions of cells with various numbers of epigenetic active alleles under different conditions. WT: wild type. $LSD1^R$: LSD1 level reduced. $E^R_{K9M}/E^{RR}_{K9M}$: H3K9 methyltransferase level reduced and further reduced. (B) Model of alleles competing for *M* enhancers. (C) Simulated allele trajectories of one cell with two epigenetically active alleles. (D) Simulated dynamics of enhancers binding to two epigenetic active alleles corresponding to the cell in panel C with the same (left) or different (by $\Delta\varepsilon = \pm 0.5\ k_BT$, middle and right) binding affinity. Also shown are schematic free energy profiles. (E) Simulated distribution of 1000 cells with various numbers of epigenetically active alleles under $E^R_{K9M}, E^{RR}_{K9M}$ and WT on day 100. (F) Fractions of overall protein expression of each allele simulated with a population of 1000 cells under $E^R_{K9M}$ and $E^{RR}_{K9M}$ comparing to those with WT. (G) The number of transcriptionally

upregulated alleles under $E^R_{K9M}, E^{RR}_{K9M}$ and in WT. (H) Schematic illustration on the mechanism of reduced OR expression diversity with $E^R_{K9M}$ and $E^{RR}_{K9M}$ compared to that in WT.

**Figure 5. Predicted OR expression switching schemes.** Typical switching examples: active pseudogene switches to intact gene (A), active intact gene switches to pseudogene and then switches to intact gene (B), and intact active gene switches off itself (C). (D) Simulated switching frequency under WT and Adcy3 KO conditions. (E) Simulated fraction of cells expressing pseudogenes under WT and Adcy3 KO conditions.

**Figure 6. The three-layer mechanism ensures mono-allele activation of OR genes.**

**Table 1 Model predictions and corresponding experimental confirmations and suggestions.** Confirmed predictions are shown as shaded. Details of suggested experiments are given in SI Materials and Methods.

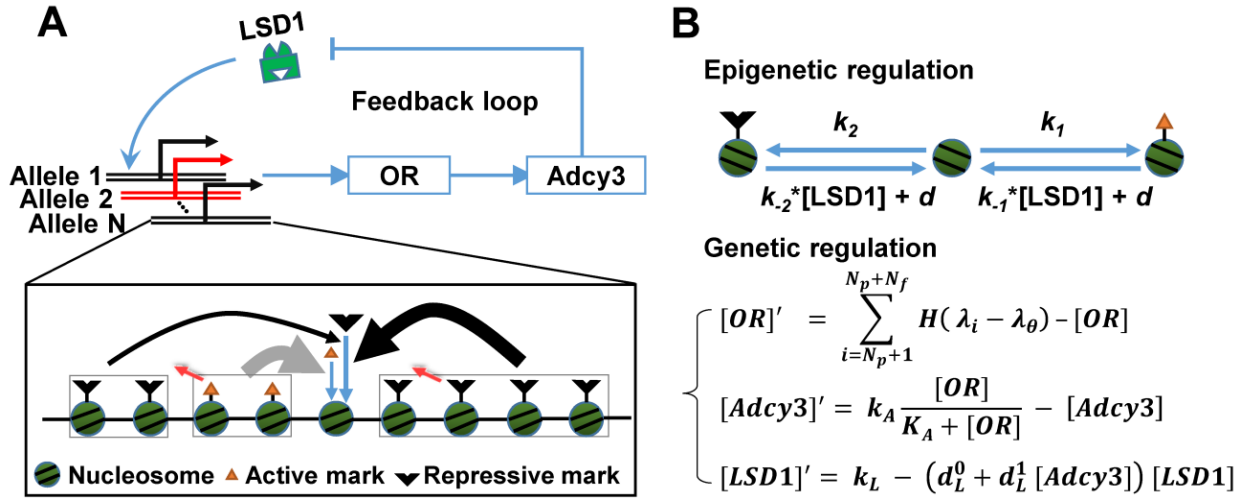

**Figure 1**

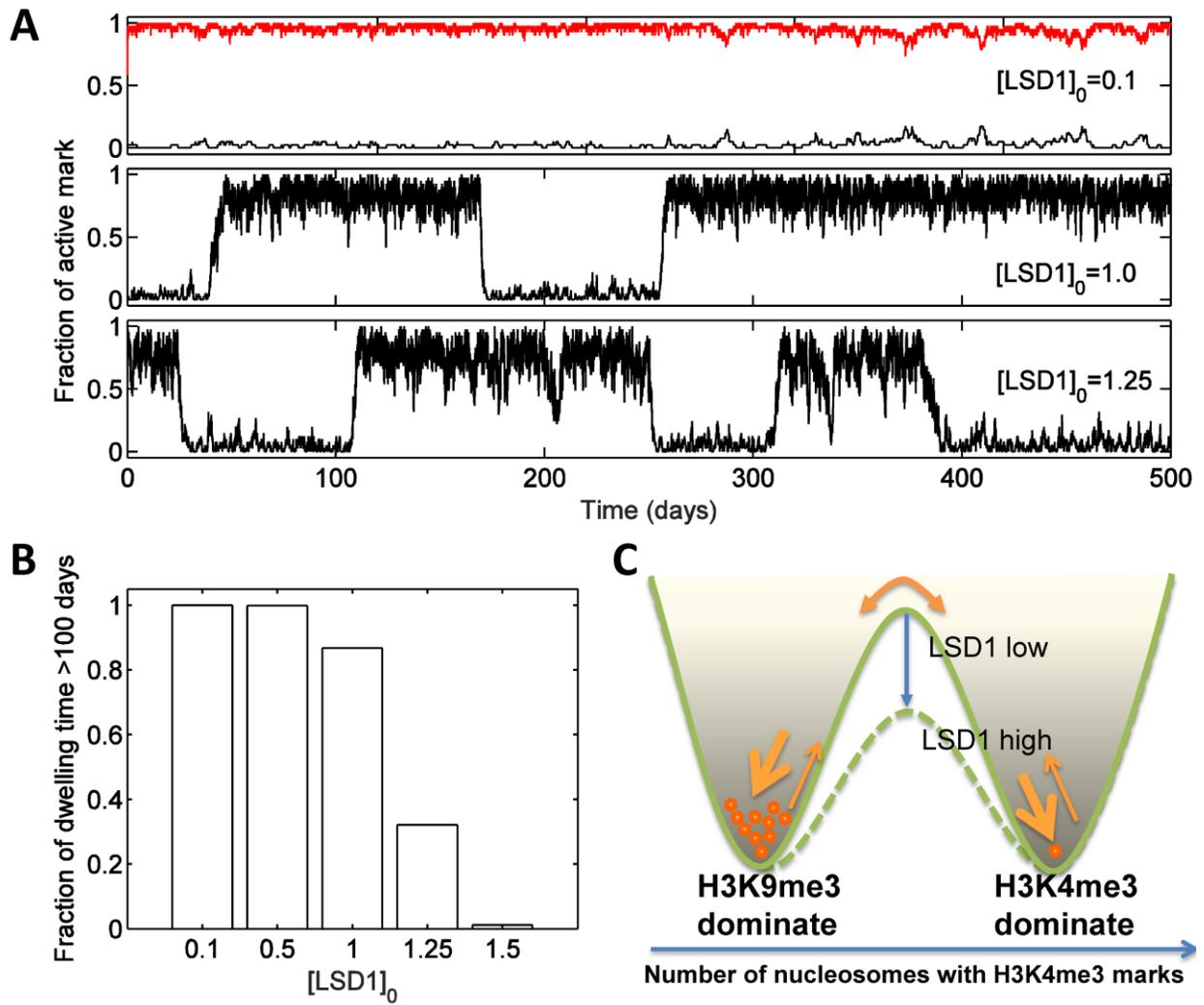

**Figure 2**

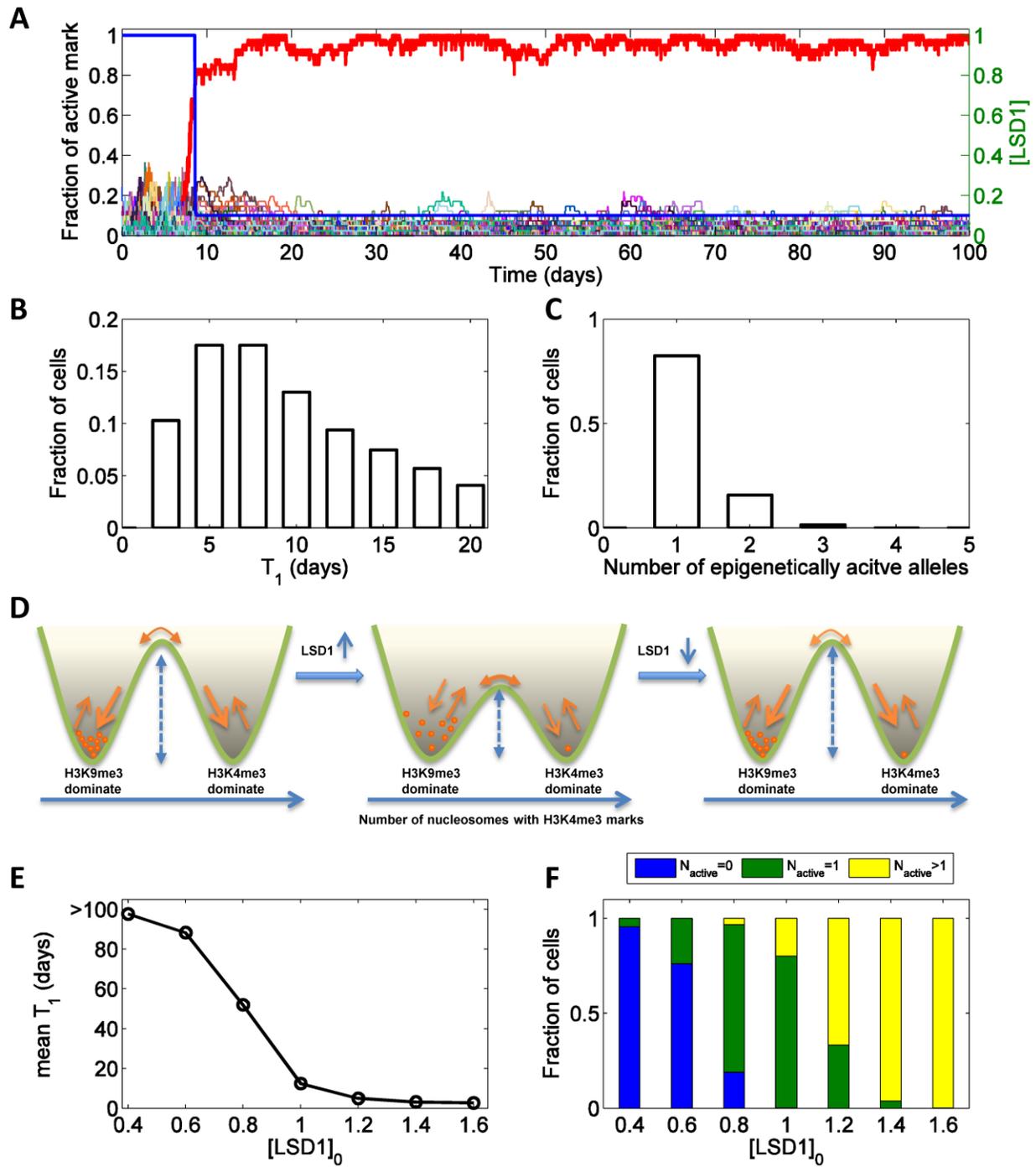

**Figure 3**

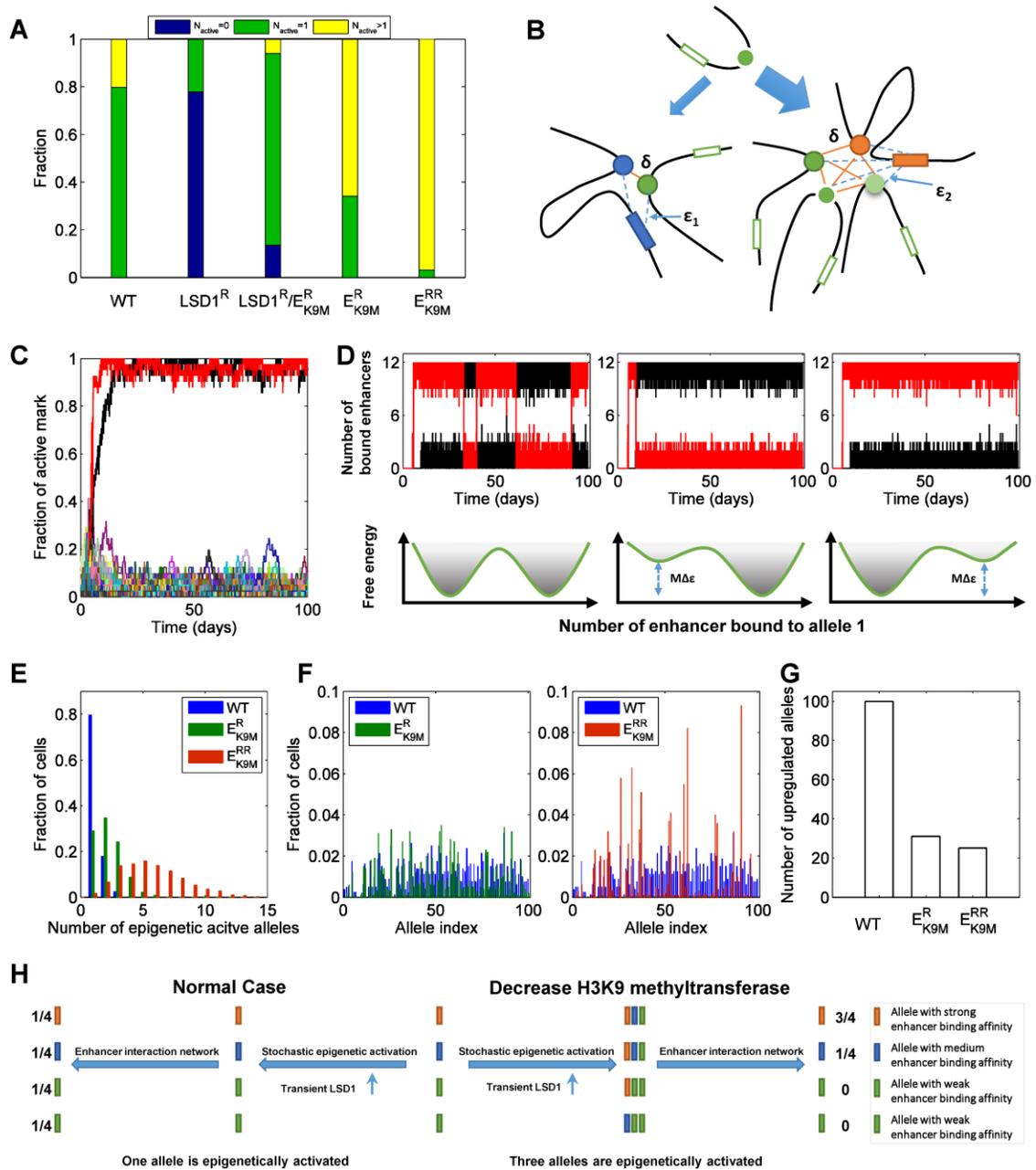

**Figure 4**

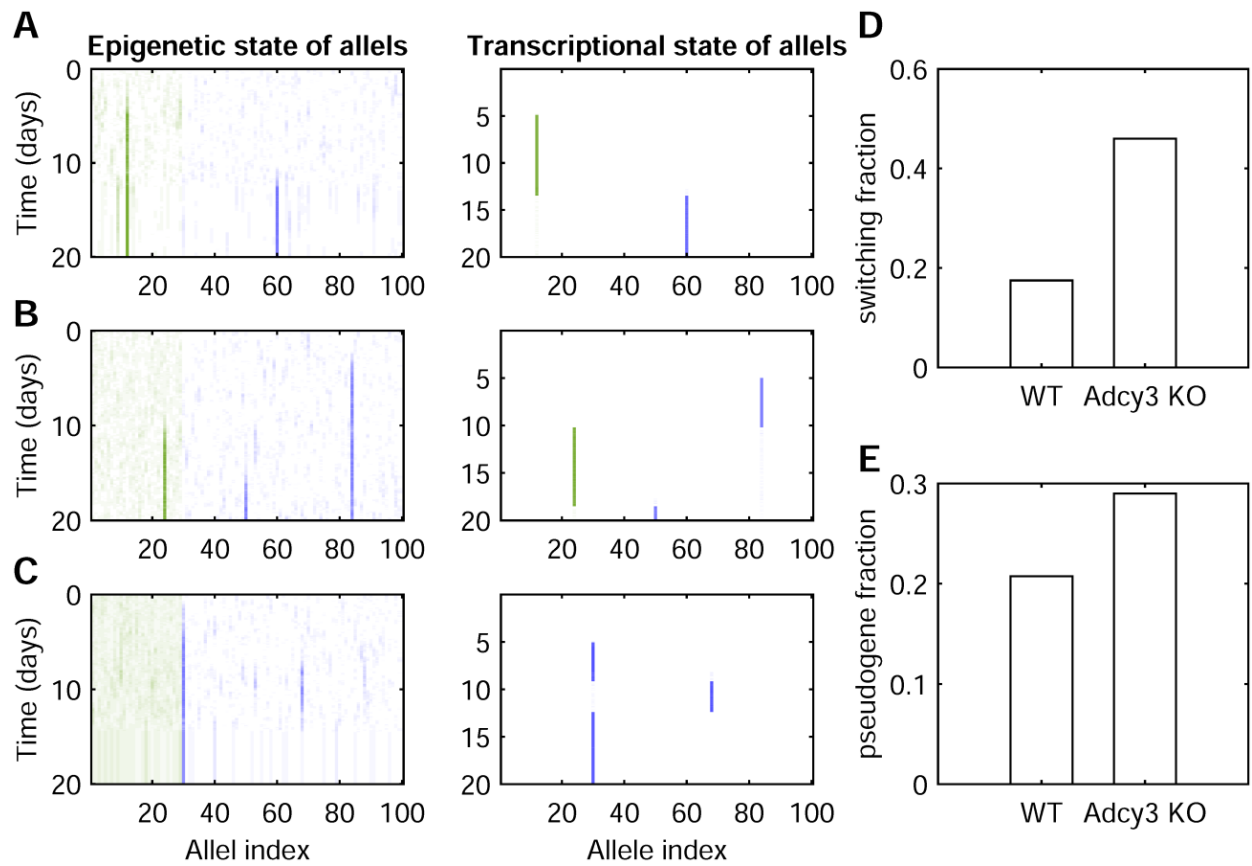

**Figure 5**

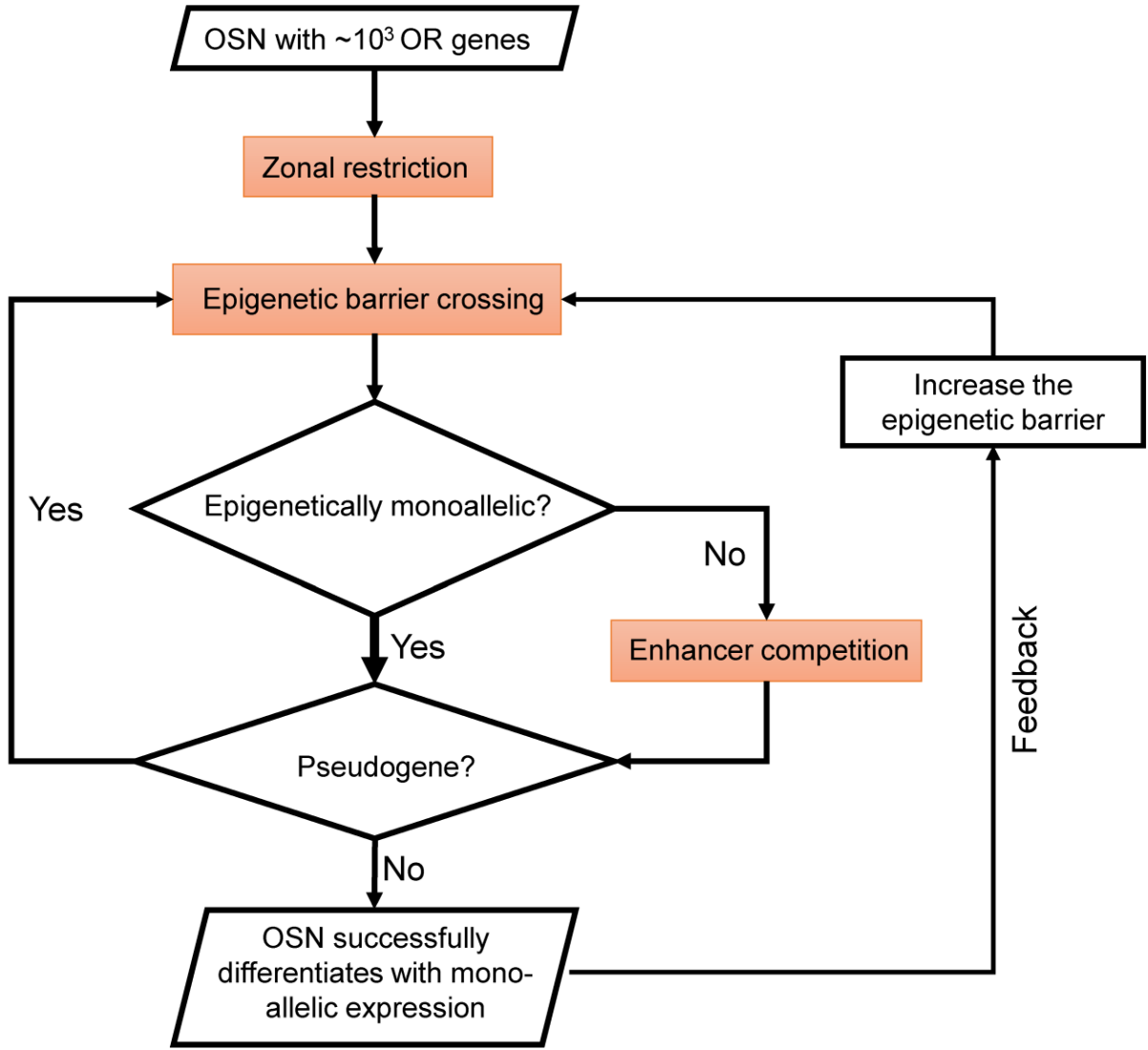

**Figure 6**

**Table 1 Model predictions and corresponding experimental confirmations and suggestions.**
Confirmed predictions are shown as shaded.

| Model predictions | Experimental confirmation or suggestions |
|---|---|
| OSNs need to maintain saturating levels of methyltransferases, but low levels of demethylases and stochastic histone exchange rate before and after differentiation (Fig. 2 and Fig. S1). | G9a/GLP at excessive concentration coexist with LSD1 during OSN differentiation (24). |
| The number of OR alleles in a zone affects the single-allele epigenetic activation ratio nonmonotonically (Fig. S2). | Introduce or remove OR alleles in a zone to test this prediction. |
| Decreasing LSD1 concentration impedes OR activation (less OSN differentiation), which can be partially restored by inhibiting G9a/GLP. | Confirmed in mice (24). |
| Epigenetic switching assumes a barrier-crossing-like dynamics for WT (Fig. 3), but a ratchet-like dynamics with G9a/GLP dKO (Fig. S4A & B). | Following Magklara et al. (20), use GFTP+ cells from OMP-IRES-GFP and G9a/GLP dKO mice, and perform CHIP-qPCR. See main text. |
| A cell may have more than one epigenetically active alleles (Fig. 3C). | Following Shykind et al. (18), cross mice bearing MOR28-IRES-Cre allele with strains bearing the reporter Rosa-loxP-stop-loxP-CFP, sort CFP$^+$Cre$^-$ cells and perform epigenetic histone modification analysis as in Magklara et al. (20). We predict that one can identify cells having the MOR28-IRES-Cre allele with H3K4me3. |
| Inhibition of H3K9 methyltransferases G9a/GLP leads to multiple allele activation (Fig. 4A). | Confirmed in Zebrafish (22) and mice (24). |
| Inhibition of G9a/GLP leads to transcriptional downregulation of most OR genes and upregulation of a small number of genes, and so decrease of diversity of expressed OR genes (Fig. 4E-F). | Confirmed in mice (24). |
| Multiple epigenetically active alleles compete for a finite number of enhancers, which contributes to the diversity reduction in G9a/GLP KO mice (Fig. 4). | Replace an upregulated OR gene and its promoter by a down-regulated one, and test whether the latter becomes upregulated in a G9a/GLP dKO main olfactory epithelium. |
| The proximity difference of enhancers to a gene leads to different OR-enhancer binding strength. | Introduce enhancers ectopically to G9a KO mice (34). We predict that the extra enhancers should at least partially rescue the reduction of OR |

|  | diversity if the number of enhancers is limited in the original G9a KO mice. |
|---|---|
| The binding strength differences between an OR promoter and individual enhancers can be small thus experimentally hard to detect, but are amplified by cooperative enhancer binding (Fig. 4D). | Lyons *et al.* did not identify any significant differences between the promoters of the most upregulated ORs and the remaining ones in predicting the transcription-factor-binding-motifs (24) |
| The switching frequency increases in Adcy3 KO OSNs compared to that in WT OSNs (Fig. 5D). Furthermore, the fraction of cells expressing pseudo ORs increases while that expressing functional ORs decreases in Adcy3 KO mice (Fig. 5E). | Confirmed (23) |
| The genes showing upregulated expression in the G9a/GLP dKO mice, such as Olfr231, have slighter stronger interactions with the enhancers than the remaining genes do. Then in normal mice, OSNs that express one of these genes should have lower frequency of switching than the cells express other genes in the same zone do. | Use techniques such as the CRISPR-Cas9 gene editing approach to fluorescently label genes like Olfr231, and perform time-lapse studies. |

# Supporting Information

**Three-layer Regulation Leads to Monoallelic and Diverse Olfactory Receptor Expression**

Xiao-Jun Tian[1,†], Hang Zhang[2†,], Jens Sannerud[3], and Jianhua Xing[1,*]

[1]Department of Computational and Systems Biology, School of Medicine, University of Pittsburgh, Pittsburgh, PA, 15260, USA

[2]Genetics, Bioinformatics, and Computational Biology Program, Virginia Polytechnic Institute and State University, Blacksburg, VA, 24060, USA

[3]The TECBio Research Experiences for Undergraduates Program, University of Pittsburg, Pittsburgh, PA, 15260, USA

[†]The two contributed equally

[*]To whom correspondence should be sent: xing1@pitt.edu

**SI Materials and Methods**

*1. Two-step procedure of reverse engineering of the feedback mechanism*

First, we focused on the requirements that after differentiation alleles with either H3K4me3 or H3K9me3 dominated states need to maintain their epigenetic state for 100 days. We asked what the constraints the requirement imposes on the model parameters, specifically on the enzyme concentrations (Fig. S1). To be general, we considered the case that the H3K9 demethylase and H3K4 demethylase are different enzymes. In this step for each set of parameters we simulated 1000 individual alleles for a duration of 100 days, and counted the percentage of alleles that maintain their collective epigenetic states. From the results in Fig. S1, we selected the following set of parameters (in reduced unit) for studies in the second step and in all other studies in the main text, $[E_1] = [E_2] = 1$, $[E_{-1}] = [E_{-2}] = 0.1$. For simplicity we assumed mass-action rate laws for all reactions. The methyltransferases in saturating concentrations (e.g., described by the Michaelis-Menten form) can further increase the robustness of the system against concentration fluctuations of these enzymes.

Second, with the final-state parameters resolved and fixed, we asked what the requirement of single allele activation within a biologically relevant time imposes on the initial enzyme concentrations prior to differentiation ($[E_1]_0$, $[E_2]_0$, $[E_{-1}]_0$ and $[E_{-2}]_0$). The change of enzyme concentrations is elicited by the negative feedback, and different ways of changing the enzyme concentrations correspond to different feedback schemes. We examined all 4 possible schemes of feedback on one enzyme and all 6 possible schemes of feedback on two enzymes. For each set of parameters we simulated 500 cells, and counted the percentage of cells that achieve single allele activation within 20 days. For each simulation, when the first allele reaches to the threshold $\lambda_\theta$, the concentration(s) of the enzyme(s) being regulated change from the initial value(s) to the final value(s). We compared this sudden switch with the continuous change modeled by ordinary differential equations as illustrated in Fig. 1 (for LSD1) and noticed no significant differences. We compared different schemes for their robustness of generating single allele activation, as well as how simple to implement the feedback (*i.e.*, the number of enzyme types to modulate).

*2. Mathematically controlled comparison of Markovian models*

We performed mathematical analysis based on the following reasoning. In the OR system a number of alleles convert their epigenetic state independently and stochastically under an elevated LSD1 concentration. Let us denote the activation time separation between the first two converted alleles as $\tau$. Then from an engineering perspective, a better design to achieve single-allele activation is the one with a larger $\tau$, which means that the two activation events are better separated temporally, and thus more time for the first allele to elicit the feedback loop and prevent activation of another allele.

Therefore we performed mathematically controlled comparison among a set of simple models shown in Fig. S4F. Consider two alleles transiting independently from the repressive-mark-dominated state to the active-mark-dominated state through (n-2) intermediate states, but with the same mean first arrival time,

$$1 \xrightarrow{k} 2 \xrightarrow{k} 3 \xrightarrow{k} ... \xrightarrow{k} n$$

Denote $p_i$ the probability of an allele in state $I$, which is given by

$$\frac{d}{dt}\begin{pmatrix} p_1 \\ p_2 \\ \vdots \\ p_n \end{pmatrix} = \begin{pmatrix} -k & 0 & 0 & 0 & 0 \\ k & -k & 0 & 0 & 0 \\ \vdots & & & & \vdots \\ 0 & 0 & 0 & k & 0 \end{pmatrix}\begin{pmatrix} p_1 \\ p_2 \\ \vdots \\ p_n \end{pmatrix}$$

$$\text{with } \begin{pmatrix} p_1 \\ p_2 \\ \vdots \\ p_n \end{pmatrix}_0 = \begin{pmatrix} 1 \\ 0 \\ \vdots \\ 0 \end{pmatrix}$$

The solution of the system is,

$$p_1(t) = e^{-kt},$$
$$p_2(t) = e^{-kt}kt,$$
$$p_3[t] \to \frac{1}{2}e^{-kt}k^2t^2,$$
$$...$$
$$p_n[t] \to \frac{1}{(n-1)!}e^{-kt}k^{n-1}t^{n-1}$$

The first-arrival time distribution is $f_n(t) = \frac{d}{dt}p_n(t)$, and fn is normalized ($\int_0^\infty f_n * dt = 1$). Then

$$f2[t] \to e^{-kt}k^1,$$
$$f3[t] \to e^{-kt}k^2t,$$
$$...$$
$$f_n[t] \to \frac{1}{(n-2)!}e^{-kt}k^{n-1}t^{n-2}$$

The mean first arrival time T is given by $T = \int_0^\infty fn * t * dt$. Requiring that the mean first arrival time T is the same for different n, one has $k = (n-1)/T$. The above formula gives the results in Fig. S4G, which shows that the two-state model has an exponentially shaped first-arrival-time distribution $f_2$, while those with ($n$ - 2) intermediate states have peaked ones that at large $t$ decrease faster with increasing $n$.

The formula below gives the distribution that the arrival time difference between two alleles is τ

$$F_n = 2\int_0^\infty f_n(t) * f_n(t+\tau) * dt$$

Thus,

$$F_n = 2\int_0^\infty \frac{1}{(n-2)!}e^{-kt}k^{n-1}t^{n-2} * \frac{1}{(n-2)!}e^{-k(t+\tau)}k^{n-1}(t+\tau)^{n-2} * dt$$
$$= 2\frac{1}{(n-2)!(n-2)!}e^{-k\tau}k^{2*n-2}\int_0^\infty e^{-2kt}t^{n-2}(t+\tau)^{n-2} * dt$$

Choose the time unit so that T = 1, one has
$$F_2 = e^{-\tau}$$
$$F_3 = e^{-2\tau}(1+2\tau)$$

$$F_4 = \frac{9}{8} e^{-3\tau}(1 + 3\tau(1 + \tau))$$
$$F_5 = \frac{1}{12} e^{-4\tau}(15 + 4\tau(15 + 8\tau(3 + 2\tau)))$$
$$F_6 = \frac{25}{384} e^{-5\tau}(21 + 5\tau(21 + 5\tau(9 + 5\tau(2 + \tau))))$$

One can randomly draw two points $p_{t1}$ and $p_{t2}$ from a distribution, corresponding to the stochastic activation events of the two independent alleles. Clearly the temporal separation of the two points, $\tau$, is likely to be larger if they are drawn from a broader $f$ corresponding to smaller $n$. Indeed Fig. S4H shows that the distribution of $\tau$ has longer tail for smaller $n$. That is, a design with the two-state dynamics is better than that with the multi-state dynamics.

**Figure S1. Low histone turnover rate stabilizes the epigenetic state of an allele.** (A) Typical trajectories of the epigenetic state of an allele with different values of the histone turnover rate. (B) Fraction of alleles that maintain epigenetic state within 100 days as a function of the histone turnover rate $d$. (C) Expanded model of histone modification reactions on a nucleosome to include methyltransferases and demethylases explicitly. (D) Fraction of alleles that maintain its epigenetic state within 100 days simulated with 256 different sets of active/repressive methylation/demethylation rates. For each parameter set 1000 independent alleles were simulated.

**Figure S2. The single-allele epigenetic activation ratio changes non-monotonically over the total number of the alleles.** Except for the total number of alleles, the simulations are performed in the same way as those in Fig. 3C. Sampled over 1000 cells at day 20.

**Figure S3. Comparative studies on all possible one-rate and two-rate feedback regulation schemes demonstrate that it is optimal to regulate both two demethylation reactions.** For each data point, the fraction of cells with one epigenetically active allele at day 20 is calculated from 500 independent simulations.

**Figure S4. Unifunctional LSD1 leads to ratchet-like dynamics and cannot ensure mono-allelic epigenetic activation.** (A) Typical trajectories of a cell show multiple-allele activation. The temporal change of LSD1 level is also indicated. (B) Distributions of the fraction of nucleosomes with active marks on day 8 show more alleles are trapped in hybrid epigenetic states with unifunctional than bifunctional LSD1. Sampled over 1000 cells. (C) The distribution of $T_1$ confirms that we chose parameters to satisfy the activation time requirement. Sampled over 1000 cells. (D) Most cells have multiple epigenetically active alleles at day 100. (E) The analogous potential system during activation illustrates the ratchet-dynamics idea (due to low H3K4 demethylation rate). (F) Minimal effective Markovian transition models for an OR allele

changing from H3K9me3 dominate state to H3K4me3 dominate state with no (n = 2, corresponding to the barrier-crossing dynamics with the bifunctional LSD1), and various number (n > 2, corresponding to the ratchet-like dynamics with the unifunctional LSD1) of intermediate states. (G) The first-arrival time (t) distribution fn of a single allele transiting from H3K9me3 dominate state to H3K4me3 dominate state as a function of the overall state number n. (H) The distribution (Fn) of first-arrival time separation ($\tau$) between two kinetically independent alleles as a function of the overall state number n. From engineering design perspective a larger $\tau$ is desirable since it gives the system more response time to elicit the feedback after the first allele becomes epigenetically active and prevents the second allele from making the transition.

**Figure S5. Enhancer competition assures transcriptional activation of single allele.** (A) Typical single-allele trajectories of the fraction of nucleosomes with active marks for 100 allele within an $E_{K9M}^{R}$ cell. (B) The distribution of the fraction of nucleosomes with active marks on day 8 averaged over 1000 cells. (C) Auxiliary enhancers stabilize binding of a specific enhancer to an allele. For each result with M enhancers, the upper one shows the trajectory of enhancer 1, and the lower one shows the corresponding number of enhancers bound to allele 1. The time is given by the number of Gillespie simulation steps. In these simulations, $\varepsilon_1 = \varepsilon_2 = -1$ $k_BT$, $\zeta = -3$ $k_BT$. Similar results were obtained with broad range of parameter values (e.g., $\varepsilon_1/\varepsilon_2$ assuming values -2 to -0.5 $k_BT$ and $\zeta$ -3 to -0.5 $k_BT$), and more enhancers involved.

**Movie S1. Illustration of the barrier-crossing like dynamics on generating mono-allelic epigenetic activation**

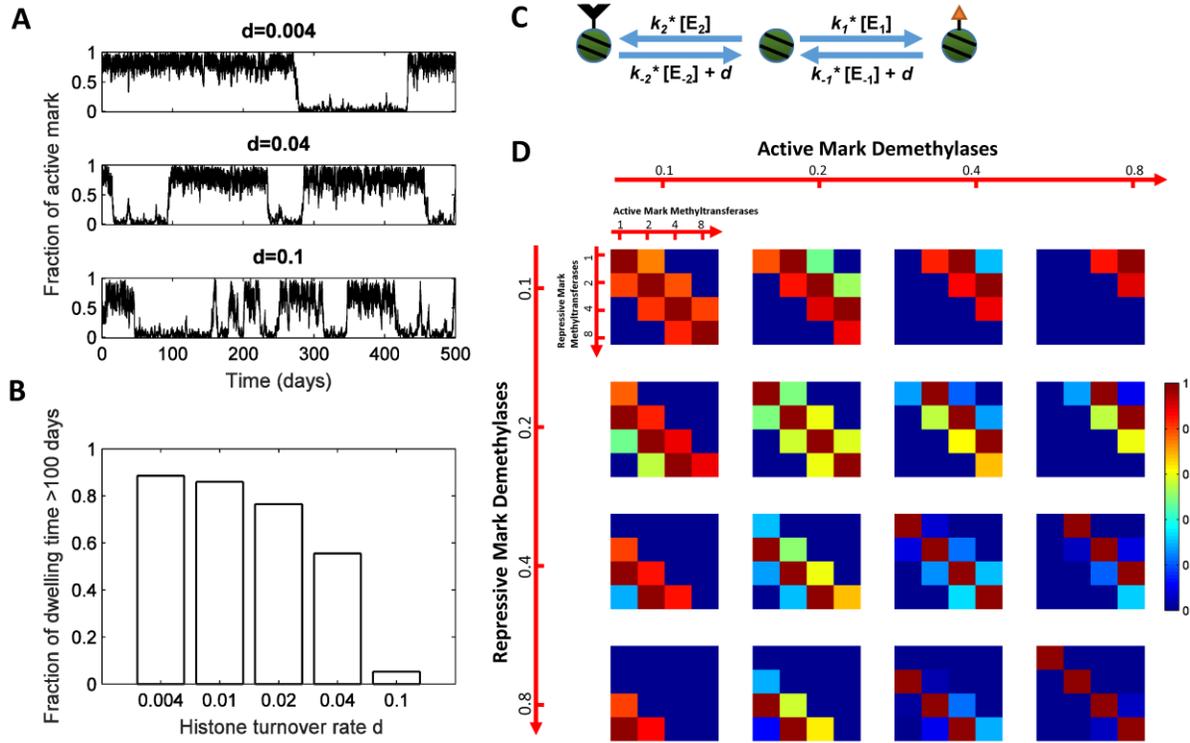

**Figure S1**

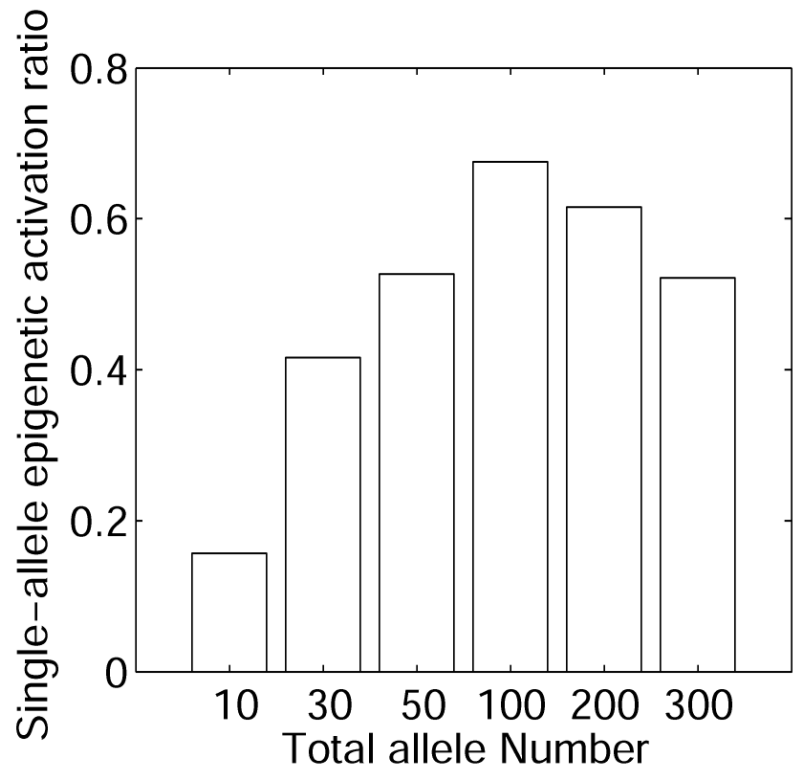

**Figure S2**

**One-Parameter Scanning**

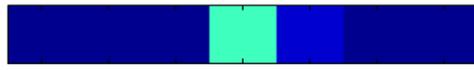
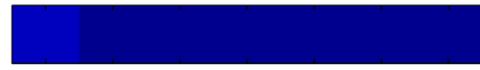
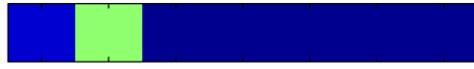
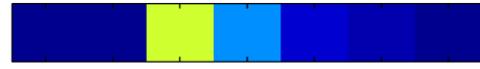

**Two-Parameter Scanning**

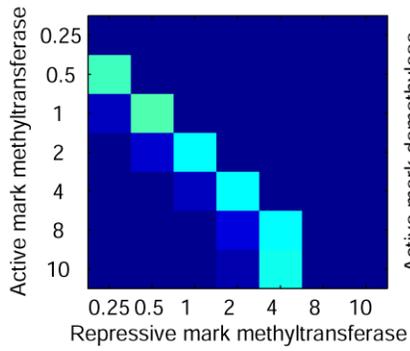
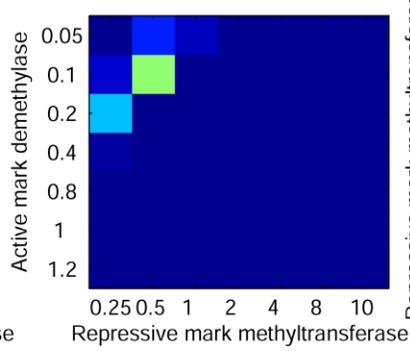
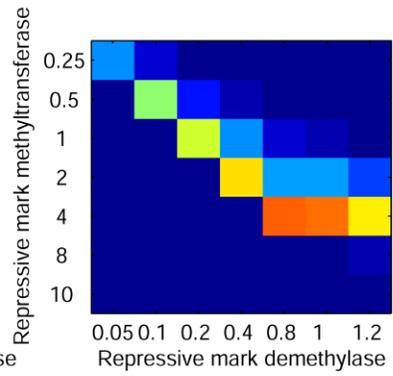
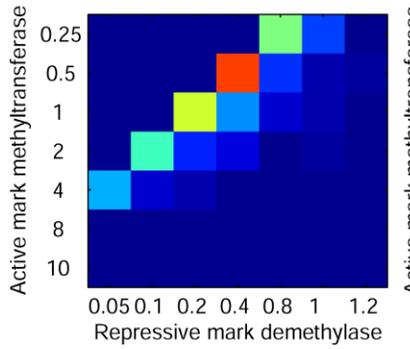
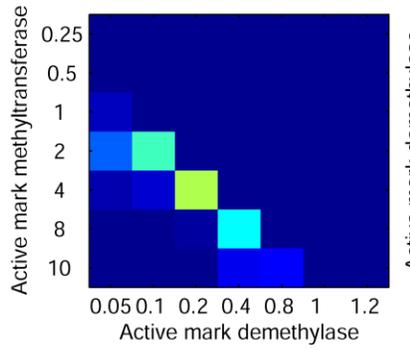
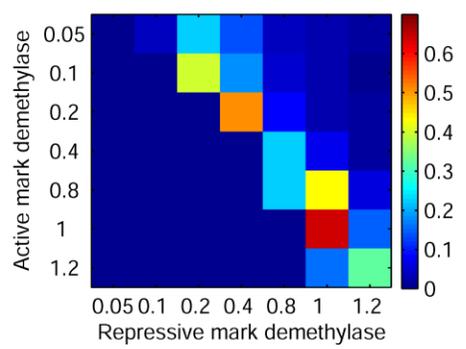

**Figure S3**

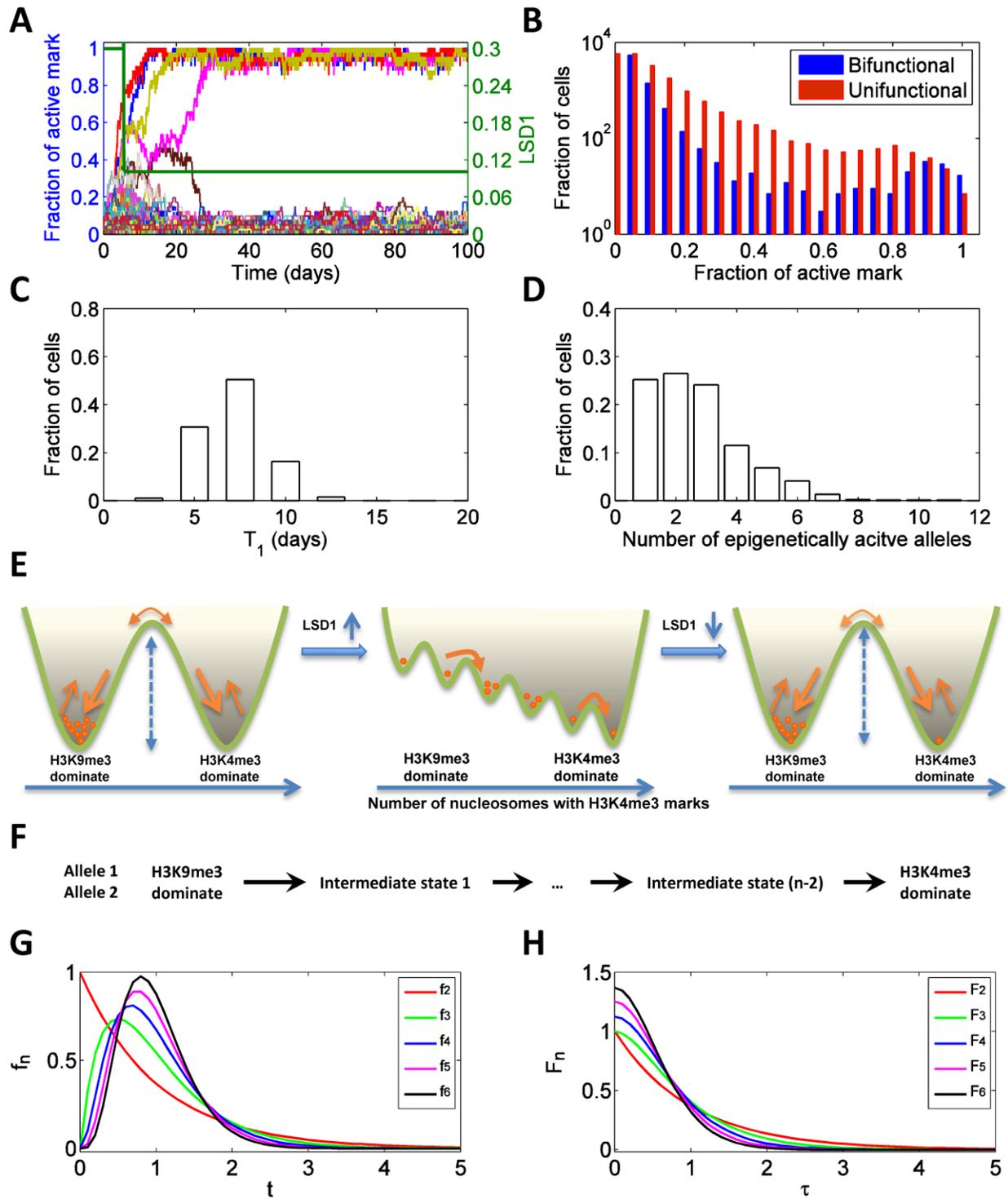

**Figure S4**

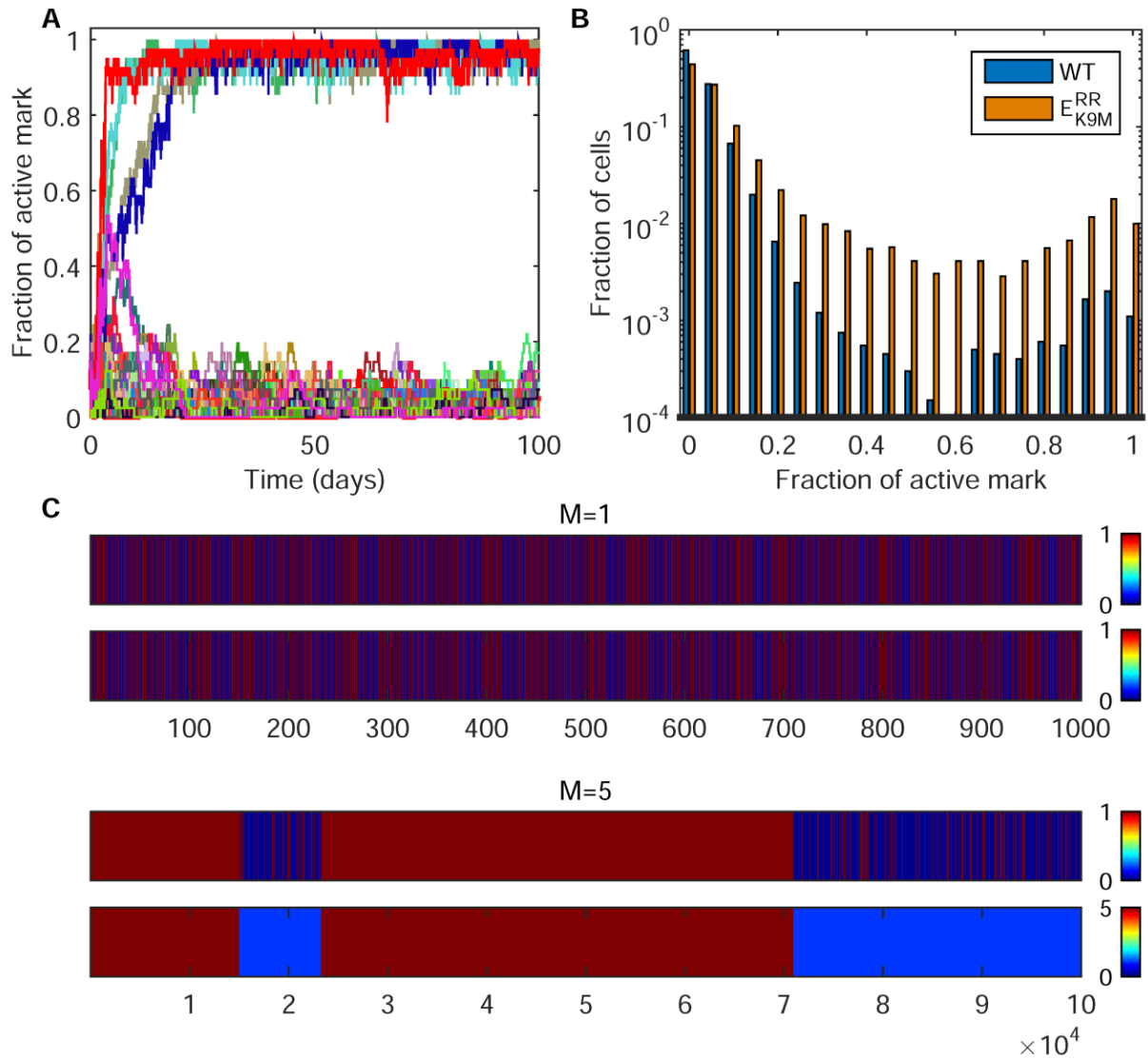

**Figure S5**

**Table S1 Values of model parameters used in this work.**

| Parameter | Value |
|---|---|
| Active mark methylation rate constant $k_1$ | within nucleation region 0.125 h$^{-1}$, outside nucleation region 0.025 h$^{-1}$. |
| Repressive mark methylation rate constant $k_2$ | within nucleation region, 0.125 h$^{-1}$ (WT), outside nucleation region, 0.025 h$^{-1}$ (WT) |
| Active mark demethylation rate constant $k_{-1}$ | 0.125 h$^{-1}$ |
| Repressive mark demethylation rate constant $k_{-2}$ | 0.125 h$^{-1}$ |
| Nucleosome correlation length $\mu$ | 0.64 |
| Histone turnover rate $d$ | 0.002 h$^{-1}$ |
| Cutoff fraction of nucleosomes with active marks so an allele is regarded as epigenetically activated $\lambda_\theta$ | 0.75 |
| Adcy3 synthesis rate $k_A$ | 1 h$^{-1}$ |
| Michaelis-Menten constant of OR induced Adcy3 expression $K_A$ | 0.8 |
| LSD1 basal degradation rate constant $d_L^0$ | 0.5 h$^{-1}$ |
| Adcy3 facilitated LSD1 degradation rate constant $d_L^1$ | 8 h$^{-1}$ |
| Prefactor for the enhancer switching rate constant $v^*$ | 1 h$^{-1}$ |
| Free energy of enhancer-enhancer interaction $\varsigma$ | -0.5 $k_B T$ |
| Free energy of enhancer-allele interaction $\varepsilon$ | ~ -1 $k_B T$ |
| Total number of enhancers $M$ | 12 |

\* This parameter is only used in generating Fig.4D for illustration purpose, and its actual value can be better estimated if time-course of OR switching becomes available.